\documentclass[aps,prl,reprint,superscriptaddress]{revtex4-1}

\usepackage{amsmath}
\usepackage{amssymb}
\usepackage{graphicx}

\pdfoutput=1

\newcommand{\eq}{\begin{equation}}
\newcommand{\eeq}{\end{equation}}
\newcommand{\ket}[1]{\left|#1\right\rangle}

\newcommand{\ohf}{\omega_\text{hf}}

\begin{document}

\title{Ultrafast Spin-Motion Entanglement and Interferometry with a Single Atom}

\author{J. Mizrahi}
\email{jmizrahi@umd.edu}
\author{C. Senko}
\author{B. Neyenhuis}
\author{K. G. Johnson}
\affiliation{Joint Quantum Institute, University of Maryland Department of Physics and National Institute of Standards and Technology,
College Park, Maryland 20742}
\author{W. C. Campbell}
\affiliation{Department of Physics and Astronomy, University of California Los Angeles, Los Angeles, CA 90095}
\author{C. W. S. Conover}
\affiliation{Colby College Physics Department, Waterville, Maine 04901}
\author{C. Monroe}
\affiliation{Joint Quantum Institute, University of Maryland Department of Physics and National Institute of Standards and Technology, College Park, Maryland 20742}

\date{\today}

\begin{abstract}
We report entanglement of a single atom's hyperfine spin state with its motional state in a timescale of less than 3 ns. We engineer a short train of intense laser pulses to impart a spin-dependent momentum transfer of $\pm 2 \hbar k$. Using pairs of momentum kicks, we create an atomic interferometer and demonstrate collapse and revival of spin coherence as the motional wavepacket is split and recombined. The revival after a pair of kicks occurs only when the second kick is delayed by an integer multiple of the harmonic trap period, a signature of entanglement and disentanglement of the spin with the motion. Such quantum control opens a new regime of ultrafast entanglement in atomic qubits.
\end{abstract}

\maketitle

Trapped atomic ions are a leading platform for quantum information processing, with a well-developed toolkit for coherent spin manipulations \cite{BlattWinelandReview}. These tools have been used to experimentally demonstrate quantum algorithms \cite{Guld,Brickman}, multiparticle entanglement \cite{Sackett,Monz}, and quantum simulations \cite{Islam,Barreiro}, among other advances. To date, most coherent manipulations of trapped ions are performed in the weak excitation regime, in which the interaction between the ions and the laser fields is characterized by a Rabi frequency $\Omega$ that is much smaller than the motional trap frequency $\omega_t$. Recent work has demonstrated coherent spin flips in the strong excitation regime, $\Omega\gg\omega_t$ \cite{Poyatos}, using picosecond laser pulses \cite{Campbell} and  near-field microwaves \cite{Ospelkaus}.  However, motional control has not been observed in the strong excitation limit.

In this Letter, we demonstrate ultrafast spin-motion entanglement, using a short train of picosecond pulses to drive stimulated Raman transitions. Each spin state receives a discrete kick in opposite directions. The momentum transfer occurs in an interaction time of 2.7 ns, only $0.2\%$ of the 1.27 $\mu$s trap oscillation period. A pair of such spin-dependent kicks, separated by an integer number of trap periods, creates an interferometer. The two spin components of the ion's wavefunction evolve along different paths in phase space after the first kick, and are then returned to their original position after the second. This is similar to other atomic interferometry experiments \cite{Kasevich,Sapiro}, with trap evolution playing the role of the atomic reflectors.

Such spin-dependent kicks are a key building block for fast entanglement of multiple ion qubits via the Coulomb interaction \cite{GarciaRipoll,Duan}. In contrast to motional gates using spectroscopically resolved sidebands, these gates may be performed faster than a trap oscillation period.  This computational speed-up comes with the additional benefits that the entangling gates will be less sensitive to noise, independent of temperature, and more easily scaled to large crystals of ions \cite{Zhu}. In addition to entangling gates, other applications of impulsive spin-dependent kicks include fast sideband cooling \cite{Machnes} and interferometry \cite{Poyatos}. 

\begin{figure}
\includegraphics{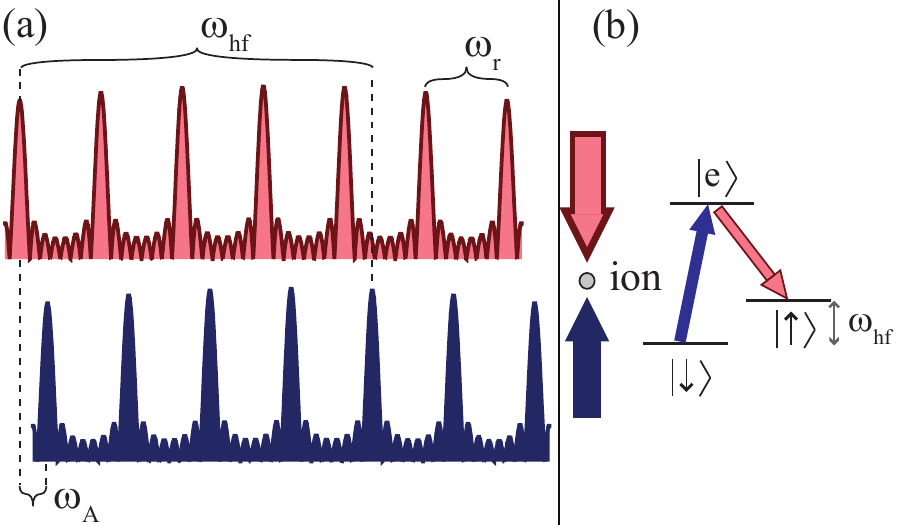}
\caption{(color online) (a): Sketch of the optical spectrum seen by the ion for generating a spin-dependent momentum kick when Eq. (\ref{eq:condition}) is satisfied. $\ohf$ is the ion's hyperfine frequency, $\omega_a$ is the AOM shift, and $\omega_r$ is the repetition frequency of the pulse train. (b): Depiction of the wavevectors associated with the spectra in (a). An atom starting in the $\ket{\downarrow}$ state may be driven to the $\ket{\uparrow}$ state only by absorbing a photon from the blue (solid) beam and emitting a photon into the red (lightly shaded) beam, resulting in a momentum transfer of $2\hbar k$ in the upward direction. Similarly, an atom starting in the $\ket{\uparrow}$ state may only make a transition such that it receives $2\hbar k$ momentum in the downward direction.}
\label{fig:combs}
\end{figure}

To create the spin-dependent kick, two pulse trains are sent onto the ion from opposite directions, with pairs of pulses from each train arriving at the ion simultaneously. The two pulse trains have a relative frequency shift between them. The ion's response can be understood in either the frequency domain or the time domain; both are instructive. First, consider the spectrum seen by the ion. As sketched in Fig. \ref{fig:combs}, the combined spectrum contains frequency components that can drive stimulated Raman transitions between the two hyperfine levels representing the effective spin $1/2$ system. However, a spin state can only undergo a transition by absorbing a photon from one beam and emitting a photon into the other beam, meaning the spin flip comes together with a momentum kick. Moreover, the kick is in opposite directions for the two spin states. In order for this process to occur, the two beams must have spectral components separated by the hyperfine frequency $\ohf$, i.e. 
\eq
\label{eq:condition}
\ohf = n\omega_{r} \pm \omega_A,
\eeq
where $n$ is an integer, $\omega_r$ is the repetition frequency of the pulse train, and $\omega_A$ is the relative frequency shift imparted by acousto-optic modulators (AOMs). The carrier-envelope phase is unimportant because the three level system driven by the comb is in a lambda configuration \cite{Hayes}. Eq. (\ref{eq:condition}) is the same condition necessary to drive a spin flip in the weak excitation limit. However, here the pulse train is nearly instantaneous compared to the trap evolution. Therefore, rather than leaving the motional state unaffected, the pulses excite all of the motional sidebands simultaneously. To avoid kicking both spin states in both directions, the width of the comb teeth must be narrower than the shift $\omega_A$, and $\ohf$ must not be an integer or half-integer multiple of $\omega_r$. The pulse train must also have sufficient intensity to drive a $\pi$ rotation.

\begin{figure}
\includegraphics{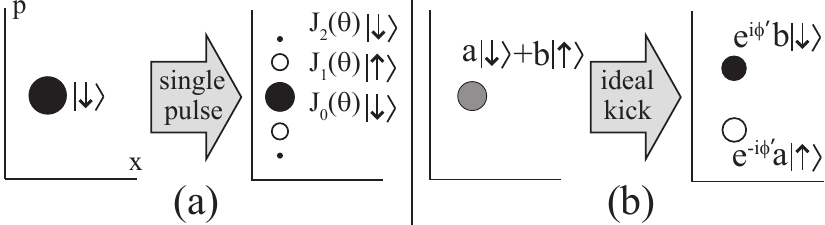}
\caption{Phase-space illustration of the different forms of spin-motion entanglement accessible in the strong-excitation regime. (a): Effect of applying a single counterpropagating pulse pair in the lin$\bot$lin polarization configuration. The initial wavepacket is diffracted into momentum states separated by $n\eta$ with amplitude $J_n(\theta)$, where $\theta$ is the pulse area, $\eta$ is the Lamb-Dicke parameter, and $n$ is an integer. A spin flip occurs for each of the odd-$n$ orders. (b): Effect of applying an ideal ``spin-dependent kick'' pulse train. The initial wavepacket is split into two momentum states entangled with the spin. }
\label{fig:pulseeffects}
\end{figure}

Alternatively, this process can be understood in the time domain as a sequence of discrete scattering pulses.  Consider the effects of a single counterpropagating pulse pair, with simultaneous arrival times and orthogonal linear polarizations.  The polarization gradient creates a standing wave in the Rabi frequency, resulting in the Hamiltonian:
\eq
\label{eq:hamiltonian}
H(t) = \omega_t a^\dagger a + \frac{\ohf}{2}\sigma_z  + \frac{\Omega(t)}{2}\sin\left[\eta(a^\dagger + a) + \phi(t)\right]\sigma_x,
\eeq
where $a$ and $a^\dagger$ are the ladder operators of the harmonic motional mode along the standing wave field, $\sigma_x$ and $\sigma_z$ are the Pauli spin operators, $\Omega(t)$ is the time-varying Rabi frequency, $\eta$ is the Lamb-Dicke parameter, and $\phi(t) = \omega_A t+\phi_0$ is the phase of the standing wave (time-dependent due to the AOM). Since the pulse is fast relative to the hyperfine and trap frequencies, we can approximate $\Omega(t) \approx  \theta \delta(t-t_0)$, and directly integrate the Hamiltonian to obtain the evolution operator for the pulse pair arriving at time $t_0$:
\eq
\label{eq:pulseoperator}
U_{t_0} = \sum_{n=-\infty}^\infty e^{i n \phi(t_0)} J_n(\theta) \mathcal{D}(i n \eta) \sigma_x^n,
\eeq
where $\mathcal{D}$ is the coherent state displacement operator. This behavior is illustrated in Fig. \ref{fig:pulseeffects}(a). The pulse creates a superposition of discrete momentum states with alternating spin states. This is Kapitza-Dirac scattering, and has previously been directly observed in atomic beams \cite{Gould}.  It is analogous to the scattering of a light wave off a thin grating, with the pulses acting as a grating to the ion's wavepacket. To create a spin-dependent kick, it is necessary to set the delays between operators of the form in Eq. (\ref{eq:pulseoperator}) such that population coherently accumulates in only the momentum orders of interest.  This is reminiscent of the temporal Talbot effect seen in matter waves \cite{Deng}, but with the various momentum states entangled with the spin.

Equation \ref{eq:pulseoperator} was derived assuming a pulse of zero duration. If $\theta$ is sufficiently small such that $J_2(\theta)\ll J_1(\theta)$ (as is the case throughout this paper), then the effect of a non-zero pulse duration will be a reduction in the effective pulse area \cite{Campbell}, which can be compensated for by increasing the laser intensity.

An analysis similar to that described in \cite{Hayes} shows that a train of such pulses may be used to generate a spin-dependent momentum kick, in which the spin states receive respective displacements in phase space of exactly $\pm i \eta$. However, unlike in \cite{Hayes}, here we need not assume the ion is in the Lamb-Dicke regime. Instead we assume a pulse train much shorter than a trap period, such that trap evolution is negligible during the pulse sequence. Under this assumption, the evolution operator for $m$ pulses arriving at times $t_1$,$t_2$,...,$t_m$ is:
\eq
\label{eq:pulsetrainoperator}
O_m = U_{t_m} e^{\frac{1}{2}i\ohf(t_m-t_{m-1})\sigma_z} \cdots U_{t_2} e^{\frac{1}{2}i\ohf(t_2-t_1)\sigma_z} U_{t_1}.
\eeq
In order for this operator to converge to a spin-dependent kick, the phases $\phi(t_i)$ imparted by each pulse and the phases imparted by free evolution must add up constructively:
\begin{align}
\Delta\phi_\text{hf} \pm \Delta\phi_\text{a} &= 2\pi n \\
\Rightarrow f_\text{hf} \pm f_\text{a} &= \frac{n}{t_k-t_{k-1}}
\label{eq:TimeDomainCondition}
\end{align}
where $\Delta\phi_\text{hf}$ is the phase advance caused by free evolution of the qubit between the pulses, and $\Delta\phi_\text{a}$ is the phase advance of the AOM phase between the pulses. For equally spaced pulses, Eq. (\ref{eq:TimeDomainCondition}) is equivalent to Eq. (\ref{eq:condition}). However, it is clear from Eq. (\ref{eq:TimeDomainCondition}) that the pulses need not be equally spaced. If the total pulse area is $\pi$ (i.e. $\theta = \pi/m$ for each pulse) and Eq. (\ref{eq:TimeDomainCondition}) is satisfied, then in the limit of many pulses Eq. (\ref{eq:pulsetrainoperator}) converges to the ideal operator,
\eq
\label{eq:sdkoperator}
U_{SDK} = e^{i\phi'}\mathcal{D}(i\eta)\sigma_\mp - e^{-i\phi'}\mathcal{D}(-i\eta)\sigma_\pm,
\eeq
where $\phi' = \phi_0 \pm m \ohf T/2$ and the signs in $\phi'$ and of the raising and lowering operators are chosen to agree with the sign in Eq. (\ref{eq:condition}). This idealized operator is shown schematically in Fig. \ref{fig:pulseeffects}(b). According to numerical simulations, with as few as eight pulses the operator in Eq. (\ref{eq:pulsetrainoperator}) can approximate the evolution described in Eq. (\ref{eq:sdkoperator}) with better than 99\% fidelity.

\begin{figure}
\includegraphics{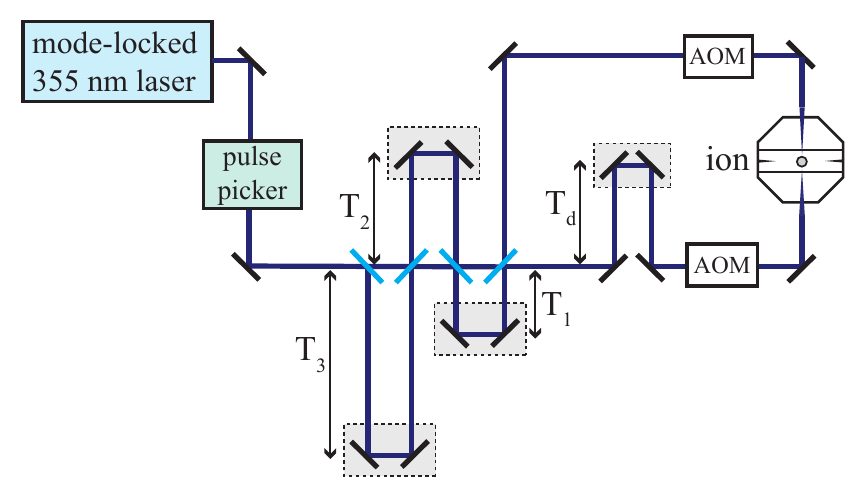}
\caption{(color online) Experimental schematic showing the pulsed laser beam path. Stacked Mach-Zehnder interferometers split each pulse into an eight-pulse sequence. Variable delays allow control of the spectrum of the eight-pulse train ($T_1$, $T_2$, $T_3$) and allow matching the arrival times of the counterpropagating pulses ($T_d$).}
\label{fig:beampath}
\end{figure}

The experimental setup is shown in Fig. \ref{fig:beampath}. A $^{171}$Yb$^+$ ion is confined in a linear four-rod Paul trap with RF drive frequency $17.9$ MHz.  The hyperfine ``clock states'' of its $^2$S$_{1/2}$ ground manifold are used as spin states, $\ket{\downarrow} \equiv \ket{F=0,m_F=0}$ and $\ket{\uparrow} \equiv \ket{F=1,m_F=0}$. These states are split by $\ohf/2\pi$ = 12.642815 GHz (including the second-order Zeeman shift from a magnetic field of 3 G).  Light near resonant with the $^2$S$_{1/2}\longleftrightarrow^2$P$_{1/2}$ transition at 369 nm is used to perform Doppler cooling, state preparation, and state detection \cite{Olmschenk}. The Raman pulse trains are derived from a picosecond mode-locked frequency-tripled vanadate laser that generates an average power of 4 W at 355 nm. This wavelength is detuned by 33 THz from the nearest excited state, and is near an optimum for minimizing spontaneous emission and differential AC Stark shifts due to the P states \cite{Campbell}. The duration of each pulse is $\tau\approx10$ ps with a repetition rate of 118.306 MHz.  An electro-optic pulse picker is used to extract individual pulses. The beam is then sent through the delay interferometers described below, frequency-shifted with AOMs, and focused onto the ion. The counterpropagating pulse trains are in the lin$\bot$lin configuration, with orthogonal linear polarizations which are also orthogonal to the quantization axis.  The beams are directed along the quantization axis (defined by the magnetic field), orthogonal to the longitudinal axis of the trap. The two radial trap frequencies are made equal, such that they are degenerate and the pulses only couple to one mode of motion. Resolved-sideband Raman spectroscopy verifies that the laser field couples mainly to a single transverse mode at $\omega_t/2\pi = 743$ kHz. The path lengths for the counterpropagating pulses are equalized to much better than $c/\tau \approx 3$ mm to match their arrival times.

As discussed above, creating a spin-dependent kick requires several pulses. However, trap evolution over the duration of even a few pulses from the laser would interfere with the production of the spin-dependent kick. It is therefore necessary to create a pulse train of shorter duration, by reshaping a single pulse into a train of pulses. This is done using concatenated Mach-Zehnder interferometers with imbalanced arm lengths. The interferometers split each pulse from the laser into a train of eight pulses with tunable relative delays, as shown in Fig. \ref{fig:beampath} (splitting the pulse further in this manner can reduce the infidelity exponentially with the number of added interferometers.)  Because the optical phase at the ion gives only a global phase shift, it need only be stable for the duration of a single experiment. Therefore, no active stabilization of the interferometers is necessary. The AOMs generate a frequency offset between the two beams of $\omega_A/2\pi = 489$ MHz. Using Eq. (\ref{eq:TimeDomainCondition}), this sets the allowable delays between each of the eight pulses to $T = 2\pi n/(\ohf+\omega_a$), where $n$ is any integer. However, we must also account for the reflective phase shift introduced by the beam splitters: pulse pairs that travel through the final delay line will have a $\pi$ phase shift relative to those that do not. To compensate for this, the final delay is set such that $n$ is a half-integer, specifically $n=5.5$ (corresponding to a delay of $T_1$ = 419 ps). Delays $T_2$ and $T_3$ are unaffected by this phase shift and are set to $n=10$ and $n=20$, respectively.  In this way, an eight-pulse train with a $2.5$ GHz repetition rate is created; it is $2.7$ ns in duration.  The pulse intensity is set such that the total area of the pulse train is $\pi$. At this intensity, a pulse train transfers an ion prepared in the $\ket{\downarrow}$ state to the $\ket{\uparrow}$ state with a fidelity of $94\%$.  This number is reduced from the theoretical maximum due to detection infidelity and micromotion (discussed below).

\begin{figure}
\includegraphics{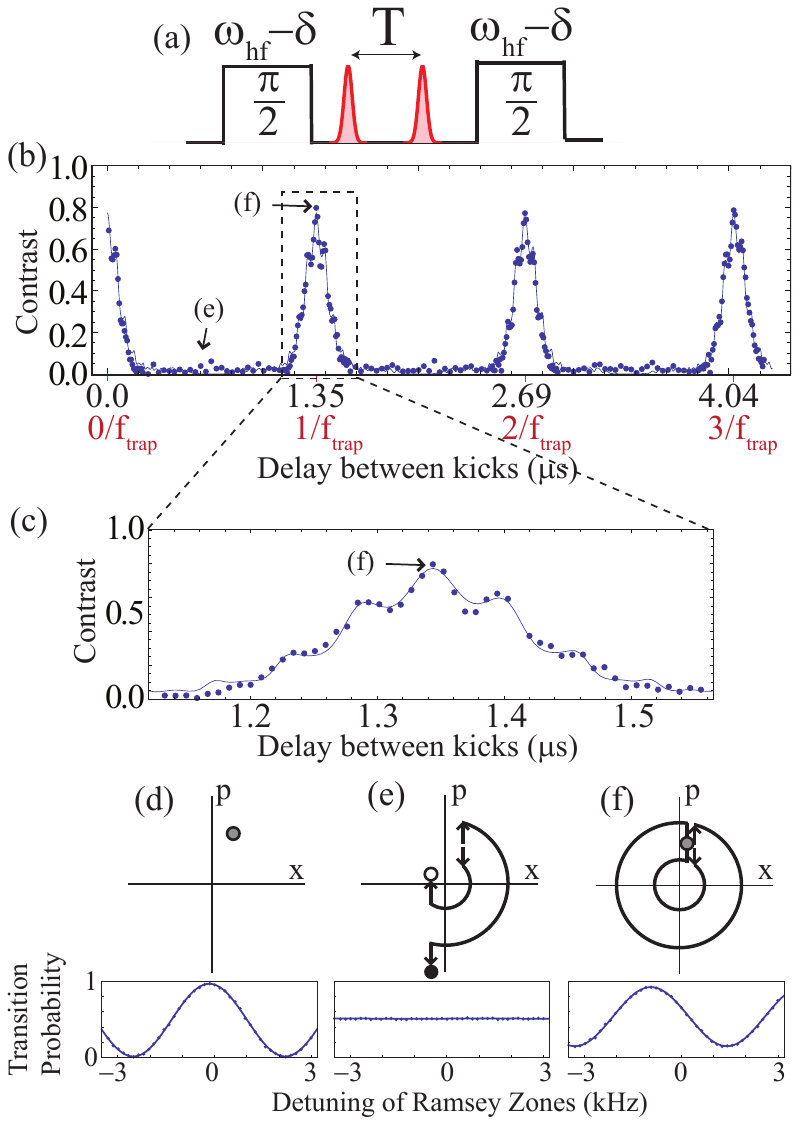}
\caption{(color online) (a): Experimental sequence to measure effect of spin-dependent kicks.  The ion is driven by two near resonant microwave $\pi/2$ pulses, with two spin-dependent kicks in between.  For each delay $T$ between the kicks, the microwave detuning $\delta$ is scanned and a Ramsey fringe contrast obtained. (b) Plot of the results of the experiment in (a). At integer multiples of the trap period, contrast revives.  (c) Close-up of the first revival peak in (b). The peak shape is a function of temperature and micromotion amplitude. The modulation of the peak is due to the better overlap of the $\ket{\uparrow}$ and $\ket{\downarrow}$ wavepackets at integer multiples of the micromotion period. The best fit curve shown is a fit to theory. Free parameters are the micromotion amplitude, average phonon number $\bar{n}$, and maximum contrast revival ($\sim$80\%). The fit shown corresponds to $\bar{n} = 10.1$. (d)-(f): Phase space plots and experimental frequency scans for different configurations.  The circle color represents spin state: black = $\ket{\downarrow}$, white = $\ket{\uparrow}$.  (e) and (f) each correspond to single points in (b), as indicated. (d) No momentum kicks; microwave pulses only.  (e) Two kicks separated by half a trap period.  (f) Two kicks separated by full trap period.}
\label{fig:contrastrevival}
\end{figure}

In order to demonstrate entanglement of the ion's spin and motion, we used a Ramsey experiment to probe collapse and revival of contrast. The Ramsey experiment is shown schematically in Fig. \ref{fig:contrastrevival}(a). The ion is initially prepared in $\ket{\downarrow}$.  It is then driven by two near resonant microwave $\pi/2$ pulses at 12.6 GHz, with a spacing between the microwave pulses of $200\:\mu \text{s}$. In between the two microwave pulses, two spin-dependent kicks are directed onto the ion.  The delay $T$ between those two kicks is then scanned.  For each value of $T$, the detuning of the microwaves $\delta$ is scanned from $-3.5$ kHz to $3.5$ kHz to obtain a Ramsey fringe. In this way, fringe contrast is measured as a function of delay between kicks. The results of this experiment are shown in Fig. \ref{fig:contrastrevival}(b).  After the first spin-dependent kick, the $\ket{\uparrow}$ and $\ket{\downarrow}$ parts of the ion's wavefunction evolve to different positions in the trap, as they now have different momenta.  In general this will destroy the spin coherence, due to the entanglement with the motion, and there will be no contrast in the Ramsey experiment.  However, at integer multiples of the trap period, the two spin states return to their original position.  Therefore, if the second spin-dependent kick arrives at an integer multiple of the trap period, it will undo the action of the first, and contrast will return.  This is shown in Fig. \ref{fig:contrastrevival}(b).  The contrast obtained with nothing between the microwaves was $97\%$ (Fig. \ref{fig:contrastrevival}(d)).  This is limited by imperfect detection.  The revived contrast at one trap period was $80\%$ (Fig. \ref{fig:contrastrevival}(f)).

The process described above is complicated by the ion's micromotion, and we find that the $\ket{\uparrow}$ and $\ket{\downarrow}$ parts of the wavefunction are better overlapped at integer multiples of the RF period than they are otherwise.  The contrast revival peak at the secular motion period is therefore modulated by smaller peaks at the micromotion period, as is seen clearly in Fig. \ref{fig:contrastrevival}(c). It is also possible that the maximum contrast of $80\%$ is due to micromotion during the pulse sequence. This effect could be suppressed by eliminating the background micromotion, or by using the axial mode of motion rather than transverse.

\begin{figure}
\includegraphics[width=\columnwidth]{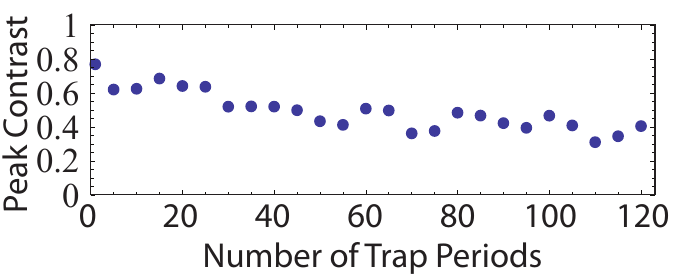}
\caption{Contrast revival for many trap revolutions.  Each point corresponds to an integer number of trap periods.}
\label{fig:manyrevivals}
\end{figure}

Fig. \ref{fig:manyrevivals} is similar to Fig. \ref{fig:contrastrevival}(b), with data points only at multiples of the trap period, showing revivals even after the ion has gone through 120 oscillations in the trap.  The slow decay here is due mainly to laser repetition rate instability, leading to timing jitter in the arrival time of the second kick relative to the first kick. This timing jitter causes a phase shift in the Ramsey fringe. Trap frequency drifts also contribute to the decay.

We have demonstrated ultrafast entanglement of an atom's spin and motion in an experimental regime that has remained largely unexplored, and used pairs of spin-dependent kicks to create an interferometer. Future work will explore the application of multiple kicks from alternating directions, increasing the area enclosed by the interferometer. In addition to improving the sensitivity of interferometric measurements, this will increase the amount of conditional phase imprinted on a pair of ions exposed to these kicks, allowing the generation of a fast controlled-phase-flip entangling gate \cite{GarciaRipoll,Duan}.

\begin{acknowledgments}
This work is supported by grants from the U.S. Army Research Office with funding from the DARPA OLE program, IARPA, and the MURI program; the NSF PIF Program; the NSF Physics Frontier Center at JQI; and the European Commission AQUTE program.
\end{acknowledgments}

\bibliography{SDKpaperrefs}

\end{document}